\documentclass[preprint, 12pt]{elsarticle}

\usepackage{svg}
\usepackage{graphicx}
\usepackage{dcolumn}
\usepackage{bm}
\usepackage{comment}
\usepackage{multirow}
\usepackage{array}
\usepackage{braket}
\usepackage{mhchem}
\usepackage{natbib}
\usepackage{tabularx}
\usepackage{amsmath, amssymb}
\allowdisplaybreaks
\usepackage{xcolor}
\usepackage[normalem]{ulem}
\usepackage{hyperref}
\usepackage{soul}
\setulcolor{red}
\usepackage{booktabs}
\usepackage{siunitx}

\usepackage{listings}
\usepackage{inconsolata}         
\definecolor{pykeyword}{rgb}{0.0, 0.5, 0.0}
\definecolor{pycomment}{rgb}{0.4, 0.4, 0.4}
\definecolor{pystring}{rgb}{0.7, 0.1, 0.1}
\definecolor{pybg}{rgb}{0.95,0.95,0.92}
\definecolor{outputtext}{rgb}{0.2, 0.2, 0.2}
\definecolor{outputbg}{rgb}{0.98, 0.98, 0.98}

\lstdefinestyle{in}{
    language=Python,
    backgroundcolor=\color{pybg},
    basicstyle=\ttfamily\small,
    keywordstyle=\color{pykeyword}\bfseries,
    commentstyle=\color{pycomment}\itshape,
    stringstyle=\color{pystring},
    showstringspaces=false,
    breaklines=true,
    frame=leftline, 
    rulecolor=\color{pykeyword},
    columns=flexible, 
    keepspaces=true
}

\lstdefinestyle{out}{
    language={},              
    backgroundcolor=\color{outputbg},
    basicstyle=\ttfamily\small
    \color{outputtext},
    breaklines=true,
    frame=none,               
    gobble=0,                  
    columns=flexible, 
  keepspaces=true
}

\renewcommand\vec{\mathbf}
\DeclareMathOperator{\tr}{tr}

\DeclareMathOperator{\St}{St}
\DeclareMathOperator{\Orb}{Orb}
\DeclareMathOperator{\proj}{Proj}

\journal{Computer Physics Communications}

\begin{document}
\begin{frontmatter}
\title{\textit{Jsymm}: A Python package for symmetry analysis of exchange tensors in magnetic Hamiltonians}

\author[label1]{Alexander S. Sergeev\corref{author}}
\author[label2]{Sergey V. Streltsov}
\cortext[author] {Corresponding author.\\\textit{E-mail address:} sergeev.physics@gmail.com}
\address[label1]{M.V. Lomonosov Moscow State University, Moscow, Russia}
\address[label2]{M.N. Mikheev Institute of Metal Physics, Ural Branch of Russian Academy of Sciences, 620137 Ekaterinburg, Russia}

\date{\today}

\begin{abstract} 
Symmetries of a crystal often restrict its physical properties. In particular, they determine possible forms of the tensors that describe interatomic exchange interaction, which governs a wide range of magnetic phenomena. Computationally demanding first-principles calculations of the exchange tensors can be greatly simplified by taking the symmetry constraints into account. Here, we present \textit{Jsymm}, a Python package that derives the most general symmetry-compatible form of the exchange tensors directly from the crystallographic data. For any bond formed by magnetic ions, \textit{Jsymm} produces the tensors of the Dzyaloshinskii--Moriya and anisotropic Heisenberg exchange interaction in symbolic form, as well as the tensors for all other bonds related to it by symmetry. This reduces the number of independent model parameters, dramatically lowering the computational cost of the \textit{ab initio} calculations and preventing unphysical results arising from symmetry violations. The package accepts standard CIF files and provides a web interface in addition to an interactive text mode and a Python library. We demonstrate its utility on La$_2$CuO$_4$ and $\alpha$-Fe$_2$O$_3$, reproducing known symmetry constraints and revealing additional relations between components of the exchange tensors of different bonds.

\noindent \textbf{PROGRAM SUMMARY}

\begin{small}
\noindent
{\em Program Title:} \textit{Jsymm}  \\
{\em Developer's repository link:} \url{https://github.com/jTraceless/jsymm} \\
{\em Licensing provisions:}  MIT License\\
{\em Programming language:} Python \\
{\em Supplementary material:} \\
{\em Nature of the problem:} A tensor describing the anisotropic exchange interaction between two magnetic moments has nine independent components. Even a relatively small unit cell of a crystal can contain tens of such pairs. Finding all their parameters in brute force \textit{ab initio} calculations is costly and inefficient. The task can be simplified by imposing symmetry constraints on the exchange tensors. However, these restrictions are to be imposed for each crystal in an \textit{ad hoc} manner. One has to identify symmetries for each bond and to find all other bonds related to it by symmetry operations, and then to compute appropriate tensors. This can become an onerous exercise in crystallography and linear algebra, which can easily lead to incomplete or incorrect results. \\
{\em Solution method:} Given a CIF file describing the structure of a crystal and a set of interatomic bonds of interest, \textit{Jsymm} first analyses the action of the crystal symmetries on the bonds, finding their stabilizer subgroups and orbits. For a representative bond in each orbit,  \textit{Jsymm} computes the symmetry-allowed exchange tensors using projection on the subspace of the trivial representation of the stabilizer subgroup in the vector space of exchange tensor matrices. Finally, \textit{Jsymm} finds the set of symmetry-compatible exchange tensors for all bonds in each orbit. This gives a set of symbolic matrices for the most general exchange tensors allowed by symmetry for the crystal. \\
{\em Additional comments:} \\
\\

\end{small}
\end{abstract}
\end{frontmatter}


\newpage
\section{Introduction}

The Heisenberg model \cite{heisenberg1928,dirac1929} has become one of the key Hamiltonians used to describe the magnetic properties of various materials with localized magnetic moments. For a pair of spins, the isotropic Heisenberg exchange interaction has the form
\begin{equation}
    H(\vec S_i, \vec S_j) = J^{\mathrm{iso}}\, \vec S_i \cdot \vec S_j = J^{\mathrm{iso}}\sum_\alpha S_i^\alpha S_j^\alpha,
\end{equation}
where $\alpha$ runs over the coordinates $x, y, z$. Although initially the coupling parameter $J^{\mathrm{iso}}$ was introduced as a scalar, it was soon realized \cite{vanvleck1937} that the exchange interaction can be rather anisotropic and should instead be described by a \emph{tensor}: 
\begin{equation}
\label{Eq:tensor-Heisenberg}
H(\vec S_i, \vec S_j) = 
  \mathbf S_i^T J \mathbf S_j
= \sum_{\alpha \beta}  S^{\alpha}_i J^{\alpha \beta}  S^{\beta}_j,
\end{equation}
where $\alpha, \beta = x, y, z$. Both the presence of the off-diagonal elements in the exchange matrix and the difference between its diagonal entries are extremely important: they determine not only the temperature and field dependence of magnetic properties, but are also responsible for various physical effects such as the inverse Dzyaloshinskii--Moriya effect (one of the key mechanisms of multiferroicity)~\cite{Cheong2007}, the formation of chiral magnetic textures, and possible realization of Kitaev spin liquids \cite{Kitaev2006,Jackeli2009}. 

While the exchange matrix is one of the key factors defining the magnetic properties of solids, its calculation using \textit{ab initio} approaches is rather challenging. Various methods have been proposed, including the Green's function technique~\cite{LEIP1, korotin2015}, the so-called four-state 
method~\cite{Xiang2011}, and approaches based on the computation of total energies for different magnetic configurations~\cite{Noodleman1981} (see also reviews \cite{Riedl2019,Li2021mol,szilva2022}). However, exchange interactions can have a long-range character and even the calculation of all isotropic exchanges $J^{\mathrm{iso}}$ can be quite time-consuming. Moreover, different methods may yield different and sometimes conflicting results.

Including spin–orbit coupling, which is necessary for computing the off-diagonal elements of the exchange tensor, substantially complicates the self-consistency procedure and slows down convergence. Together with on-site Coulomb repulsion (Hubbard $U$) it also leads to the formation of local minima of energy functional, often causing the calculations to become trapped in such states and leading to incorrect results as a consequence. Thus, evaluating all elements of the exchange tensor, even for a limited set of nearest neighbors, becomes a remarkably complex problem.

The number of parameters to be calculated can be substantially reduced if the symmetry of the system is taken into account \cite{izyumov1991,tsukerblat1983}. In many cases, knowledge of the point group  allows one to restrict the problem to the calculation of just a few matrix elements of the exchange tensor. In this work, we present \textit{Jsymm}, a Python package that allows one to find the form of symmetry-compatible exchange tensors for any set of bonds in a crystal of interest, using only crystallographic information from a \verb|cif| file (such files are available in public databases). The article is organized as follows. We start by describing a mathematical algorithm for finding the symmetry constraints in Sec.~\ref{Sec:SymCons}. In Sec.~\ref{Sec:ImpDet} we discuss the details of how the algorithm is implemented. Sec.~\ref{Sec:RunPack} provides examples of working with the package. Finally, we demonstrate the results of the symmetry analysis for several physical examples in Sec.~\ref{Sec:Test}, and conclude in Sec.~\ref{Sec:Sum}.

\section{Symmetry constraints on exchange tensors}\label{Sec:SymCons}
Consider a bond consisting of two magnetic ions with exchange interaction described by Eq.~\eqref{Eq:tensor-Heisenberg}. We decompose the exchange tensor into symmetric and antisymmetric parts:
\begin{equation}
\label{Eq:Hamiltonian}
H( \vec S_i, \vec S_j) =  
\vec S_i^T \Gamma  \vec S_j
+ \vec D \cdot [\vec S_i \times \vec S_j].
\end{equation}
Here, the symmetric part is given by $\Gamma = \frac 12 (J + J^T )$ and the antisymmetric part is represented by a three-dimensional vector $\vec D$ describing the Dzyaloshinskii--Moriya interaction (DMI). Although the symmetric part includes the isotropic exchange, $J^{\mathrm {iso}} = \frac 13 \tr (\Gamma)$, we will refer to $\Gamma$ as the symmetric anisotropic exchange in what follows. 

In a crystal, symmetry operations can interchange atoms and transform their spins. Since the full Hamiltonian must be invariant under this action, the symmetries can severely restrict the possible form of the exchange matrices. The symmetry constraints come in two forms:
\begin{itemize}
    \item The tensors of each individual bond must be invariant under the action of the symmetry elements that map the bond to itself.
    \item  If a symmetry element maps one bond into another, their tensors must be compatible with this action. 
\end{itemize}
The first kind of constraints affects only the anisotropic exchange tensors, $\vec D$ and the traceless part of $\Gamma$. For the DM exchange interaction, these constraints are known as the ``Moriya rules'' \cite{Moriya1960}. But even if such constraints are absent, there can be non-trivial relations of the second kind. In particular, the isotropic exchange $J^{\mathrm {iso}}$ must have the same value for all bonds related by symmetry operations.  

Our goal is to determine for a given set of bonds the most general form of $\vec{D}$ and $\Gamma$ allowed by crystal symmetry. To this end, we first examine the action of symmetry elements on the exchange matrices, then consider the action on the bonds in the context of the periodic lattice, and finally determine the desired invariant form of exchange matrices both for an individual bond and for all other bonds related to it by the symmetry operations.

\subsection{Symmetry element action on exchange tensors}
To find the constraints imposed by symmetry on the exchange tensors of a bond, we first need to understand how these tensors are transformed under the action of symmetry elements. Recall that spins transform as axial vectors:
\begin{equation}\label{Eq:gS}
g \cdot \vec S = \det(R_g) R_g \vec S,
\end{equation}
where $g$ is a symmetry element, $R_g$ is its matrix, and $\vec S = (S_x, S_y, S_z)^T$. We define the corresponding action on the Hamiltonian term by
\begin{equation}\label{Eq:gH}
[g \cdot H] (\vec S_i, \vec S_j) = H(g^{-1} \vec S_i, g^{-1} \vec S_j).
\end{equation}

An exchange term in the Hamiltonian can be written as
\begin{equation}\label{Eq:ExchMatrix}
H_J(\vec S_i, \vec S_j) = \vec S_i^T J \vec S_j,
\end{equation}
where $J$ is the exchange matrix. In particular, the DMI matrix has the form
\begin{equation}
\label{eq:D}
D = \begin{pmatrix}
0 && D_z && -D_y \\
-D_z && 0 && D_x \\
D_y && -D_x && 0 \\
\end{pmatrix},
\end{equation} 
where $D_\alpha$ are the components of the vector $\vec D$. Since the matrix $D$ is antisymmetric, we have
\begin{equation}
\vec S_i^T D \vec S_j = - \vec S_j^T D \vec S_i,
\end{equation} 
which one can interpret as
\begin{equation}\label{Eq:DSign}
H_D(\vec S_i, \vec S_j) = H_{-D} (\vec S_j, \vec S_i).
\end{equation}
In other words, the vector $\vec D$ must change sign upon the permutation of the spin labels to compensate the sign change of the cross product and leave the energy contribution invariant.

Suppose that the group element $g$ maps each atom of the bond to itself. Then it follows from Eqs. \eqref{Eq:gS}, \eqref{Eq:gH} and \eqref{Eq:ExchMatrix} that  the matrix $J$ transforms as
\begin{equation}
g\cdot J = R_g J R_g^T,
\end{equation}  
where we used that $R_g$ is orthogonal, so $R_{g^{-1}} = R_g^{-1} = R_g ^T$. In the case when $g$ interchanges two atoms, the DMI matrix also changes its sign, according to Eq.~\eqref{Eq:DSign}. We conclude that the symmetry element that maps the bond to itself acts on the exchange matrices as follows:
\begin{equation}\label{Eq:gGD}
g\cdot \Gamma = R_g \Gamma R_g^T, \qquad
g\cdot D = \pm R_g D R_g^T,
\end{equation} 
where the minus sign corresponds to the symmetries that flip the bond.

One can also derive the transformation rule for the vector $\vec D$. Since the cross product transforms as an axial vector, 
\begin{multline}\label{Eq:DTransform}
[g\cdot H_{\vec D}](\vec S_i, \vec S_j) = 
\vec D \cdot [R_g^T \vec S_i \times R_g^T\vec S_j] = \\
\vec D \cdot \bigl(\det(R_g^T) R_g^T [\vec S_i \times \vec S_j] \bigr) = 
\bigl(\det(R_g) R_g  \vec D \bigr) \cdot [\vec S_i \times \vec S_j].
\end{multline}
Thus, $\vec D$ also transforms as an axial vector, which additionally changes its sign when $g$ flips the bond.

\subsection{Stabilizer and orbit of the bond}\label{Sec:StabOrbBond}
Our next step is to determine which symmetry elements contribute to the constraints of the first kind introduced above and which bonds are affected by the constraints of the second kind. To this end, we define the notions of a stabilizer subgroup and an orbit of a bond, taking into account the lattice periodicity. 

Recall that the point group $G$ of a crystal is defined as a quotient $G = S/T$ of the space group $S$ by the subgroup $T$ of lattice translations. The point group consists of cosets
\begin{equation}
[g] = \{R| \vec t\}T,
\end{equation}  
where $R$ is the reflection or rotation matrix and $\vec t$ is the translation vector. For each coset $[g]$, we choose a representative element $g = \{R| \vec t\}$, where $\vec t$ has non-negative fractional components. 

Our goal is to define the action of the point group $G$ on the bonds formed by pairs of atoms. As an intermediate step, consider the action of the representative $g$ on an atom $a$ with the position vector $\vec r$:
\begin{equation}\label{Eq:ga}
g\cdot a = \{\text{atom at position } g\cdot  \vec r = R\vec r + \vec t\}.
\end{equation} 
We define a bond as an ordered pair of atoms and denote it 
\begin{equation}
\vec b = (a_1, a_2).
\end{equation}
There is no need to distinguish bonds that are related by a uniform lattice translation of both atoms. Formally, one can describe this by an equivalence relation $\sim$ on the set of all bonds:
\begin{equation}\label{Eq:EqTr}
(a_1 + \vec T, a_2 +\vec T) \sim (a_1, a_2), 
\end{equation}
where $\vec T$ is a translation by a lattice vector. Now we can consider the equivalence classes of bonds $[\vec b]$ associated with $\sim$. One can think of the class $[\vec b]$ as an infinite lattice of all bonds related to $\vec b$ by a lattice translation. 

We define the action of the point group on the class $[\vec b] = [(a_1, a_2)]$ as
\begin{equation}\label{Eq:gb}
[g] \cdot [\vec b] = [(g\cdot a_1, g\cdot a_2)].
\end{equation}
One checks that the action is well-defined in both arguments: the result does not depend neither on the choice of the representative $g$ of the coset, nor on the choice of the bond $\vec b$ in the equivalence class. This happens because the lattice translations in $[g] = gT$ are absorbed by the equivalence of bonds. Note, however, that it is crucial to ``mod out'' the lattice translations only as  a last step, not for the symmetry action on individual atoms. Keeping this in mind, from now on we will not distinguish between the coset $[g]$ and its representative $g$, to simplify the notation.

The action \eqref{Eq:gb} allows us to introduce the standard notions of a stabilizer and an orbit. Note that the equivalence \eqref{Eq:EqTr} distinguishes the bonds that have the opposite orientation: $\vec b = (a_1, a_2)$ and $-\vec b = (a_2, a_1)$ belong to different classes. However, an important constraint on the exchange tensors of the bond comes from the symmetries that flip the bond. Taking this into account, we define the stabilizer of the bond $[\vec b]$ as
\begin{equation}
\St = \{g\in G \mid g\cdot [\vec b] =[\pm \vec b]\}. \label{Eq:Stab}
\end{equation}
In words, the stabilizer subgroup consists of the symmetry elements that, up to a lattice translation, either leave the bond fixed or interchange its atoms. This set of symmetry operations determines constraints on the form of exchange tensors for a given bond class. The orbit
\begin{equation}
\Orb = \{ g\cdot [\vec b] \mid g\in G\}. \label{Eq:Orb}
\end{equation}
contains all the bonds whose exchange tensors are symmetry-related to those of the starting bond. Note that, for the orbit-stabilizer theorem to hold, the elements of the orbit should also be considered ``modulo orientation'' to avoid double-counting. 

\subsection{Symmetry-compatible exchange matrices}\label{Sec:SymMat}
To find the most general form of the exchange matrices for the bonds in the class $[\vec b]$ and its orbit, we use basic tools from the representation theory of finite groups. Since the DMI matrix $D$, Eq.~\eqref{eq:D}, has three independent matrix elements, one can consider it as an element of a three-dimensional vector space $V_D$:
\begin{equation}
D = \sum_{\alpha = x, y, z} D_{\alpha} e_{\alpha}.
\end{equation} 
Here, $e_{\alpha}$ are the basis elements in the space of antisymmetric rank 3 matrices. For example,
\begin{equation}
e_x =\begin{pmatrix}
0 && 0&& 0\\
0&& 0&& 1\\
0 && -1 &&0 \\
\end{pmatrix}.
\end{equation}
In a similar way, the anisotropic symmetric matrix $\Gamma$ belongs to the six-dimensional vector space $V_{\Gamma}$ of rank 3 symmetric matrices. Since the action \eqref{Eq:gGD} preserves (anti)symmetry of the exchange matrix $J$, it can be interpreted as a linear transformation of the corresponding vector space $V_{J}$:
\begin{equation}\label{Eq:RepRho}
(g\cdot J)_i = \sum_j \rho_{ij}(g) J_j,
\end{equation}
where $J_j$ is a component of a $d$-dimensional vector of matrix elements ($d = 3$ for $J=D$ and $d = 6$ for $J = \Gamma$).  The set of such linear transformations 
\begin{equation}
\rho = \{\rho(g) \mid g\in \St \}
\end{equation}
forms a representation of the stabilizer $\St$ of the bond class $[\vec b]$, which is a subgroup of the point group $G$. 

An exchange matrix $J$ for the bond in the class $[\vec b]$ is symmetry-compatible if 
for all elements in the stabilizer we have
\begin{equation}
g\cdot J = J, \qquad g\in \St.
\end{equation} 
This means that $J$ belongs to the invariant subspace of the trivial representation $W_1 \subset V_J$. Thus, our task of finding all symmetry-compatible matrices $J$ amounts to specifying the subspace $W_1$. First, we determine its dimension:
\begin{equation}\label{Eq:dimW}
\dim W_1 = \frac{1}{|\St|}\sum_{g\in \St} \tr \rho(g).
\end{equation}  
If $\dim W_1=0$, the anisotropic exchange interaction $J$ is not allowed by symmetry.
If $\dim W_1 = d$, that is, the whole space transforms according to the trivial representation, then the stabilizer of the bond does not put any symmetry constraints on the matrix $J$ (but there may be symmetry relations between the matrices for different bonds in the orbit).

In the intermediate case $0< \dim W_1 < d$  we need to find $n = \dim W_1$ linearly independent vectors in the subspace $W_1$. This is done by the projection operator
\begin{equation}\label{Eq:ProjW}
\proj_1 = \frac{1}{|\St|} \sum_{g\in \St} \rho(g),
\end{equation}
which projects any vector onto the invariant subspace $W_1$ of the trivial representation. Applying $\proj_1$ to the basis elements $\{e_i\}$, one can find the desired basis for the space $W_1$. 

Once the subspace $W_1$ is found for the bond class  $[\vec b]$, one can compute symmetry-compatible exchange matrices for all bonds in the orbit of $[\vec b]$ using  the requirement that $H$ be equal to $g\cdot H$ for all pairs of sites. Note that the bond $[\vec b]$ may be mapped to $g\cdot [\vec b]$ by more than one group element. One checks that the resulting Hamiltonian $g\cdot H$ does not depend on the choice of this element, by virtue of using $g^{-1}$ in Eq.~\eqref{Eq:gH}. 

For a specific example, let $g$ map the bond $(a_1,a_2)$ to the bond $(a_3, a_4)$. Then the invariance $g\cdot H = H$ implies that 
\begin{equation}\label{Eq:J34}
J_{34} = R_g J_{12} R_g^T
\end{equation} 
for each exchange matrix.

\section{Implementation details}\label{Sec:ImpDet}
Below we discuss general features of data structures and algorithms, which we use to implement the mathematical procedures described in the previous section.
\subsection{Crystal data}
Information on chemical composition, cell vectors, and coordinates of atoms is obtained from a \verb|cif| file, which is processed using the \verb|ase| library \cite{HjorthLarsen2017}. We also check that the site occupancy is an integer for all atoms with the \verb|parsnip-cif| package \cite{parsnip-cif}.

Then the list of atomic coordinates is processed with the \verb|spglib| library \cite{Togo2024}, which finds the space group of the crystal and provides the list of all symmetry elements. Each symmetry element consists of the matrix part and the translation part, both expressed in the basis of the cell vectors. The only control parameter for the symmetry analysis is the numerical precision \verb|sym_tolerance|, which is used to determine whether coordinates of atoms coincide after applying symmetry operations. Its default value is $10^{-5}$~\AA. Note that this parameter might influence the resulting symmetry group in some cases.

For each symmetry element, we compute the Cartesian version of the matrix part. To this end, we construct the basis transformation matrix from the cell vectors. After the transformation, we convert the resulting matrix to a \verb|SymPy| matrix object, whose elements are exact expressions, such as $\tfrac{\sqrt 3}{2}$. The list of possible exact values is determined by the fact that a crystal can only have rotation axes of orders $1, 2, 3, 4$ and $6$. We assume that the crystal structure is defined using a conventional unit cell in the standard crystallographic setting. Then the matrix elements can only take values $0$, $\pm 1$, $\pm \tfrac{1}{2}$ and $\pm \tfrac{\sqrt{3}}{2}$. 

All geometric data, including coordinates of atoms and symmetry elements, is stored in an instance of class \verb|Compound|. 

\subsection{Representation of atoms and bonds}
Atoms are represented by instances of \verb|AtomicSite| class. An atom is specified by an index $n$, which enumerates atoms in the unit cell, and by the integer cell coordinates $(m_x, m_y, m_z)$. The value of $n$ encodes both the fractional part of atom's scaled coordinates $\vec r_f$, which take value in the interval $[0, 1)$, and its chemical element. We will refer to the cell with coordinates $(0, 0, 0)$ as the \emph{home unit cell}.

We define the \verb|Bond| object as an ordered pair of \verb|AtomicSite| objects. Since we are interested in bonds only up to a lattice translation, we define the \emph{standard set} as a subset of the set of all possible bonds. The bond belongs to the standard set if the following condition is satisfied: in each pair of cell coordinates, $(m_x^1, m_x^2), (m_y^1, m_y^2), (m_z^1, m_z^2)$, the smallest coordinate is $0$. In other words, the bonds in the standard set belong to the sector with non-negative cell coordinates and lie as close to the origin as possible.
In the language of Sec. \ref{Sec:StabOrbBond}, the standard set is a set of representatives for equivalence classes that correspond to the equivalence relation \eqref{Eq:EqTr}. Moving the bond to the standard set after any transformation  allows us to work with the equivalence classes $[\vec b]$.

\subsection{Action of symmetry elements}
Symmetry elements are represented by instances of \verb|SymElement| class, which contain rotation/reflection matrices $R$ and translation vectors $\vec t$ in scaled and Cartesian coordinates. 

\subsubsection{Action on atoms}
Schematically, the action \eqref{Eq:ga} of a symmetry element $g = \{R| \vec t\}$ on an atom is implemented as follows: 
\begin{equation}
\vec r_f + \vec m \quad = \quad 
\vec r \quad \xrightarrow{g} \quad  
R\vec r + \vec t \quad  = \quad   
 \vec r' \quad  \xrightarrow{\text{dec.}} \quad
 \vec r'_f + \vec m'.
\end{equation}
Here, all vectors and symmetry transformations are expressed in the scaled coordinates. First, we add fractional and integer coordinates of an atom, then act on the resulting vector with the symmetry transformation, and then decompose the result into fractional and integer parts. In the last step, one must carefully handle the cases when atom is close to the cell boundary, which is done as follows for each coordinate:
\begin{enumerate}
\item If a coordinate lies within \verb|sym_tolerance| of an integer $m$, round it to $m$.
\item Use Python \verb|floor| function to extract the integer part, to be interpreted as a cell coordinate (recall that \verb|floor(x)| returns the largest integer not greater than \verb|x|).
\item Interpret the remaining fractional part as the coordinate inside the home unit cell.
\end{enumerate}
Finally, we determine the atomic index by comparing the fractional coordinates with those of the atoms in the home unit cell. 

\subsubsection{Action on bonds}
The action of a symmetry element on a bond consists of two steps, which implement  Eq.~\eqref{Eq:gb}:
\begin{enumerate}
\item Act on each atom individually.
\item If necessary, move the resulting bond to the standard set by parallel translation.
\end{enumerate} 

In the context of orbits, each \verb|Bond| object can play one of two roles: it can be a \emph{starting bond} of the orbit, or an element of an orbit of another bond. Consider a starting bond $\vec b$, which is represented by an object \verb|b|. The \verb|orbit| property of \verb|b| is a list of bonds of the form $g\cdot \vec b$, each containing the link to the starting bond \verb|b| and the index of the symmetry element \verb|i_g|. The \verb|orbit| is constructed in such a way that all its elements are distinct, so there is some arbitrariness in choosing the symmetry element $g$. The full information about group action is stored in the \verb|full_orbit| property, which contains all bonds of the form $g\cdot \vec b$ with their respective group elements.

\subsection{Finding symmetry constraints on exchange tensors}
An exchange tensor is represented by an instance of \verb|ExchangeTensor| class. Its properties include $3\times 3$ \verb|SymPy| matrix and a list of symbols used in the matrix. There are two types of exchange tensors, \verb|Symm| and \verb|DMI|, whose matrices are symmetric and antisymmetric, respectively. Each tensor can be represented as a ``vector'' of independent matrix elements. For \verb|DMI| tensor, this vector is $(D_x, D_y, D_z)$, while for the symmetric exchange it reads $(\Gamma_{xx}, \Gamma_{yy}, \Gamma_{zz}, \Gamma_{xy}, \Gamma_{xz}, \Gamma_{yz})$. 
\subsubsection{Constraints for individual bond}
The method \verb|compute_exchange()| of a \verb|Bond| object  finds the most general form of exchange tensors allowed by symmetry for the given bond. 

First, we find the matrices of representation $\rho(g)$, as defined by Eq.~\eqref{Eq:RepRho}. To this end, we compute the action defined by Eq.\eqref{Eq:gGD} using Cartesian rotation matrices for $R_g$. Then, according to Eq.~\eqref{Eq:RepRho}, we consider the $i$-th component of the transformed vector as a linear combination of initial vector components. To find the coefficients, we employ \verb|SymPy| methods for polynomials:
\begin{equation}
\rho_{ij} \leftrightarrow \verb|sp.Poly(g_J_v[i], J.symb).eval(vals) |
\end{equation}
 For example, for the DMI exchange tensor we treat $(g\cdot D)_i \leftrightarrow \verb|g_J_v[i]|$ as a polynomial in variables $(D_x, D_y, D_z) \leftrightarrow \verb|J.symb|$, and evaluate it for 
\begin{equation}
\{D_j = 1, \text{  other components are zero}\} \leftrightarrow \verb|vals|,
\end{equation} 
which gives the desired coefficients $\rho_{ij}(g)$.

Then, proceeding with the steps described in Sec.~\ref{Sec:SymMat}, we find the dimension of the invariant subspace of the trivial representation \eqref{Eq:dimW}. If needed, we construct the projection operator \eqref{Eq:ProjW} and act with it on basis vectors of the form $\vec e_i = (0, \ldots, 0, 1, 0, \ldots, 0)^T$ with non-zero $i$-th component. We check that the result is non-zero, normalize it, and check that it is linearly-independent with the previously found vectors by using the $\verb|nullspace()|$ method of $\verb|SymPy Matrix|$. The process continues until we find $n = \dim W_1$ vectors, which span the invariant subspace. Finally, we multiply each $\proj_1 \vec e_i$ by the $i$-th symbol (e.g. $D_i$) and convert the resulting vector of components into matrix form, which gives the symmetry-adapted exchange tensor matrix.

\subsubsection{Symmetry relations between bonds in the orbit}
\label{Sec:SymRelOrb}
Suppose that we have computed exchange tensors for a bond $\vec b $. Then we can find the exchange tensors for the bond $g\cdot \vec b$ in the orbit according to Eq.~\eqref{Eq:J34}. This is implemented in the \verb|pull_exchange()| method of the class \verb|Bond|. One should \emph{not} use \verb|compute_exchange()| for this purpose, since this method treats the bond individually, and its symmetry relations with other bonds in the orbit are lost.

A typical algorithm for finding the symmetry-compatible exchange tensors for all bonds in an orbit consists of two steps:
\begin{enumerate}
\item Choose a starting bond $\vec b$ and compute its exchange tensors by \\ \verb|compute_exchange()|.
\item Iterate over the bonds in the orbit of $\vec b$ and  call \verb|pull_exchange()| for each bond.
\end{enumerate}
This will produce a set of exchange tensors that is invariant under the action of the symmetry group.

\subsection{Building list of bonds}
Our code provides functionality for a common task: for a given crystal, find all symmetry constraints on the exchange tensors for bonds between magnetic elements $A$ and $B$, whose length does not exceed a given value $l$. 

We start by listing all atoms and their positions for a $3 \times 3 \times 3$ cluster of cells centered at the home unit cell. Then we use the \verb|cKDTree| method of the \verb|SciPy spatial| module to build a list of pairs of atoms $a_A, a_B$ such that 
\begin{equation}
\text{distance}(a_A, a_B) < l.
\end{equation} 
The first atom always lies in the home unit cell, while the second atom can be anywhere in the cluster. Next, we create \verb|Bond| objects from all pairs, forming a pool of bonds. Our goal is to organize these bonds by their symmetry and length. The result will be the list of starting bonds, each being the first in its orbit. The orbit of each starting bond will be stored as an attribute of the \verb|Bond| object. This is done as follows:
\begin{enumerate}
\item Pick any bond $\vec b$ from the pool and compute its symmetry orbit. Compute the exchange tensors for all bonds in the orbit. Add $\vec b$ to the list of bonds.
\item Exclude the orbit elements from the pool of bonds.
\item Repeat until the pool is empty. 
\end{enumerate}
Finally, we sort the bonds in the list by their length.

\section{Running package}\label{Sec:RunPack}

In this section, we briefly demonstrate how one can use the package in various modes. We start with an interactive text mode, then consider the graphical web interface, and finally discuss using the package as a code library. For more usage examples and a detailed description of the package API, we refer the interested reader to the documentation \cite{jsymm-docs}.  

Note that the code's output has a dual nature: on the one hand, it is strictly determined by  symmetry, but on the other hand, it can assume seemingly different forms. When analyzing the results, one should keep in mind that the essential part of the output is the relations between the matrix elements of the exchange tensors for bonds in one orbit. In particular, each independent variable has the same value in all these matrices (of course, there is no such relationship between the matrix elements for bonds in different orbits). 

Two possible sources of variation in the form of the exchange matrices are:
\begin{itemize}
    \item Arbitrary choice of orientation of bonds that enter the orbit, which influence the sign of the DMI tensor.
    \item Degree of alignment between the starting bond of an orbit and the coordinate axes of the Cartesian coordinate system. Better alignment leads to simpler form of the matrix elements. 
\end{itemize}

\subsection{Interactive text mode}\label{Sec:IntMode}
Given a \verb|cif| file of a crystal, one can quickly analyze the exchange tensors using the package in a \verb|Python| console. Below we show a typical workflow, using hematite $\alpha$-Fe$_2$O$_3$ as an example.

We start by initializing the \verb|Compound| object from a \verb|cif| file. 
\begin{lstlisting}[style = in]
$ python -m jsymm
=== Welcome to jsymm interactive session ===
Available classes: Compound, Bond, BondList

>>>Fe2O3 = Compound("Fe2O3.cif")
>>>Fe2O3.show()
\end{lstlisting}

\begin{lstlisting}[style = out]
Symmetry group: R-3c (167),  |G| = 36
Number of atoms: 30
[Fe: 12,  O: 18]
\end{lstlisting}
The output indicates that the \verb|Compound| was initialized successfully. 

Now one can find all bonds between Fe atoms, whose length does not exceed 5\AA:
\begin{lstlisting}[style = in]
>>>BL = BondList(Fe2O3, "Fe", "Fe", 5)
>>>BL.show()
\end{lstlisting}
\begin{lstlisting}[style = out]
List of starting bonds:
ind                 Bond               L, A    |Orb|   DMI     Symm
0. (Fe_8, [0 0 1])--(Fe_6, [0 0 0])     2.8873  6       1       2
6. (Fe_7, [0 0 0])--(Fe_12,[0 0 0])     2.9671  18      0       6
24. (Fe_10,[0 0 0])--(Fe_8, [0 0 0])    3.3651  18      2       4
42. (Fe_7, [0 1 0])--(Fe_10,[0 0 0])    3.7013  36      3       6
78. (Fe_7, [0 0 0])--(Fe_8, [0 0 0])    3.9872  6       0       2
\end{lstlisting}
Here, an atom is represented by a label, which indicates the atom's position in the unit cell, and a triple of integer unit cell coordinates. The list contains starting bonds of the orbits. Also shown are the length of each bond and the number of bonds in its orbit. The columns \verb|DMI| and \verb|Symm| show the numbers of independent matrix elements in the exchange matrices $D$ and $\Gamma$. One can readily conclude that the DMI exchange in the second and in the last orbit is prohibited by symmetry. 

Exchange tensors for the bonds from the first orbit are
\begin{lstlisting}[style = in]
>>>BL.show_ex_orbit(0)
\end{lstlisting}
\begin{lstlisting}[style = out]
Orbit of the bond 0. (Fe_8, [0 0 1])--(Fe_6, [0 0 0]) and DMI exchange tensor components:

ind		g * b				 [x, y, z]
0. 	(Fe_8, [0 0 1])--(Fe_6, [0 0 0])   [0, 0, Dz]
1. 	(Fe_11,[0 0 0])--(Fe_9, [0 0 1])   [0, 0, Dz]
2. 	(Fe_10,[0 0 0])--(Fe_12,[0 0 0])   [0, 0, -Dz]
3. 	(Fe_5, [0 0 0])--(Fe_7, [0 0 0])   [0, 0, -Dz]
4. 	(Fe_3, [0 0 0])--(Fe_1, [0 0 0])   [0, 0, Dz]
5. 	(Fe_2, [0 0 0])--(Fe_4, [0 0 0])   [0, 0, -Dz]

Orbit of the bond 0. (Fe_8, [0 0 1])--(Fe_6, [0 0 0]) and Symm exchange tensor components:

ind		g * b				 [xx, yy, zz, xy, xz, yz]
0. 	(Fe_8, [0 0 1])--(Fe_6, [0 0 0])   [Gxx, Gxx, Gzz, 0, 0, 0]
1. 	(Fe_11,[0 0 0])--(Fe_9, [0 0 1])   [Gxx, Gxx, Gzz, 0, 0, 0]
2. 	(Fe_10,[0 0 0])--(Fe_12,[0 0 0])   [Gxx, Gxx, Gzz, 0, 0, 0]
3. 	(Fe_5, [0 0 0])--(Fe_7, [0 0 0])   [Gxx, Gxx, Gzz, 0, 0, 0]
4. 	(Fe_3, [0 0 0])--(Fe_1, [0 0 0])   [Gxx, Gxx, Gzz, 0, 0, 0]
5. 	(Fe_2, [0 0 0])--(Fe_4, [0 0 0])   [Gxx, Gxx, Gzz, 0, 0, 0]
\end{lstlisting}
The first table indicates that the DMI vector of the bonds belongs to the vertical axis. To interpret the signs correctly, one should take into account the orientation of the bonds:
\begin{lstlisting}[style = in]
>>>BL.show_orbit(0)
\end{lstlisting}
\begin{lstlisting}[style = out]
Orbit of the bond b = (Fe_8, [0 0 1])--(Fe_6, [0 0 0]):
|St| = 6,	|Orb| = 6,	L = 2.8873 A
dim DMI = 1	dim Symm = 2

ind	g			g * b			Bond vector, A
0. 	g_0 (Fe_8, [0 0 1])--(Fe_6, [0 0 0])  [0.0000, 0.0000, -2.8873]
1. 	g_1 (Fe_11,[0 0 0])--(Fe_9, [0 0 1])  [0.0000, 0.0000, 2.8873]
2. 	g_6 (Fe_10,[0 0 0])--(Fe_12,[0 0 0])  [0.0000, 0.0000, 2.8873]
3. 	g_7 (Fe_5, [0 0 0])--(Fe_7, [0 0 0])  [0.0000, 0.0000, -2.8873]
4. 	g_13 (Fe_3, [0 0 0])--(Fe_1, [0 0 0])  [0.0000, 0.0000, 2.8873]
5. 	g_18 (Fe_2, [0 0 0])--(Fe_4, [0 0 0])  [0.0000, 0.0000, 2.8873]
\end{lstlisting}
To orient the bonds uniformly, we need to flip the first and fourth bonds, which will change the signs of $\vec D$. 

Now consider the fourth orbit. It contains 36 bonds, which equals the order of the group. This means that each of these bonds is stabilized only by the identity transformation. Still, symmetry puts constraints on the exchange matrices:
\begin{lstlisting}[style = in]
>>>BL.show_ex_orbit(42, 'DMI')
\end{lstlisting}
\begin{lstlisting}[style = out]
Orbit of the bond 42. (Fe_7, [0 1 0])--(Fe_10,[0 0 0]) and DMI exchange tensor components:

ind             g * b                            [x, y, z]
42.     (Fe_7, [0 1 0])--(Fe_10,[0 0 0])         [Dx, Dy, Dz]
43.     (Fe_12,[0 0 0])--(Fe_5, [0 1 0])         [Dx, Dy, Dz]
44.     (Fe_7, [0 0 0])--(Fe_10,[1 0 0])         [-Dx/2 - sqrt(3)*Dy/2, sqrt(3)*Dx/2 - Dy/2, Dz]
45.     (Fe_12,[1 0 0])--(Fe_5, [0 0 0])         [-Dx/2 - sqrt(3)*Dy/2, sqrt(3)*Dx/2 - Dy/2, Dz]
46.     (Fe_7, [0 0 0])--(Fe_10,[0 0 0])         [-Dx/2 + sqrt(3)*Dy/2, -sqrt(3)*Dx/2 - Dy/2, Dz]
47.     (Fe_12,[0 0 0])--(Fe_5, [0 0 0])         [-Dx/2 + sqrt(3)*Dy/2, -sqrt(3)*Dx/2 - Dy/2, Dz]
48.     (Fe_9, [1 0 0])--(Fe_8, [0 0 0])         [-Dx/2 + sqrt(3)*Dy/2, sqrt(3)*Dx/2 + Dy/2, -Dz]
49.     (Fe_6, [0 0 0])--(Fe_11,[1 0 0])         [-Dx/2 + sqrt(3)*Dy/2, sqrt(3)*Dx/2 + Dy/2, -Dz]
50.     (Fe_9, [0 0 0])--(Fe_8, [0 1 0])         [Dx, -Dy, -Dz]
<...>
\end{lstlisting}
Thus, while each DMI vector is described by three independent variables, the vectors are not at all arbitrary, since they are related by the symmetries that interchange the bonds in the orbit.

\subsection{Web interface}
The package provides a web interface that can be run locally \cite{jsymm-docs} or accessed on the web \cite{jsymm-web}. The user can upload a \verb|cif| file and analyze it with a chosen numerical precision. Alternatively, one can load an example file from the Help page. Once the file is loaded, the interface shows crystal geometric data, such as position of atoms and representation of symmetry elements. The user can then choose a pair of chemical elements and a maximal bond length $L$, and obtain a list of all bonds shorter than $L$ classified into symmetry orbits. 

For a chosen orbit, the interface shows  a table of symmetry-compatible exchange tensors. For any bond, the user can further address its geometric details (coordinates of atoms), its stabilizer subgroup, and interact with a 3D model showing  the placement of the bond inside the unit cell.

\subsection{Usage as a library}\label{Sec:UseLib}
\subsubsection{Orbit of a bond}
Consider a bond $\vec b$ in La$_2$CuO$_4$ compound. The orbit of the bond is simply a list of distinct bonds obtained from the starting bond by the group action. Each of these bonds, $g\cdot \vec b$, stores the starting bond $\vec b$ (as a property \verb|starting_bond|) and the index of the group element $g$ (as a property \verb|i_g|):
\begin{lstlisting}[style = in]
from jsymm import *
LCO = Compound("La2CuO4-Sol.cif")
b = Bond(LCO).from_labels("Cu_4", [0,0,0], "Cu_1", [0,0,0])
for gb in b.orbit:
    print(f"g_{gb.i_g} * b = ", gb)
\end{lstlisting}
\begin{lstlisting}[style = out]
g_0 * b =  (Cu_4, [0 0 0])--(Cu_1, [0 0 0])
g_1 * b =  (Cu_4, [0 1 0])--(Cu_1, [1 0 0])
g_2 * b =  (Cu_1, [0 0 0])--(Cu_4, [0 1 0])
g_3 * b =  (Cu_1, [1 0 0])--(Cu_4, [0 0 0])
g_8 * b =  (Cu_2, [1 0 0])--(Cu_3, [0 0 0])
g_9 * b =  (Cu_2, [0 1 0])--(Cu_3, [0 0 0])
g_10 * b =  (Cu_3, [0 0 0])--(Cu_2, [1 1 0])
g_11 * b =  (Cu_3, [0 0 0])--(Cu_2, [0 0 0])
\end{lstlisting}
The stabilizer of the bond is a dictionary in which the keys are the indices of the symmetry elements that map the bond to itself, and each value is of \verb|Boolean| type.  It equals \verb|True| if the symmetry element flips the bond and \verb|False| otherwise:
\begin{lstlisting}[style = in]
print(b.stabilizer)
\end{lstlisting}
\begin{lstlisting}[style = out]
{0: False, 6: True}
\end{lstlisting}

\subsubsection{Computing exchange}
The exchange matrices for an individual bond can be found by the method \verb|compute_exchange()|. The results of the  computation are stored as a dictionary with keys \verb|"DMI"| and \verb|"Symm"|, and the values are instances of the class \verb|ExchangeTensor|. Each tensor can be represented as a column vector or as a $3\times 3$ matrix, and also contains the list of all symbols that can enter its matrix. Using these symbols, one can evaluate the exchange tensor for specific numerical values of the variables:
\begin{lstlisting}[style = in]
b.compute_exchange()
print(b.exchange['DMI'], "\n")
Dx, Dy, Dz = b.exchange['DMI'].symb
print(b.exchange['DMI'].evaluate({Dy: 3, Dx: 1}))
\end{lstlisting}
\begin{lstlisting}[style = out]
[Dx, Dy, 0] 

[[ 0  0 -3]
 [ 0  0  1]
 [ 3 -1  0]]
\end{lstlisting}
The numbers of independent matrix elements are stored in the attribute \verb|dimV|:
\begin{lstlisting}[style = in]
for ex_type, dim in b.dimV.items():
    print(f"dim {ex_type} = {dim}")
\end{lstlisting}
\begin{lstlisting}[style = out]
dim DMI = 2
dim Symm = 4
\end{lstlisting}
If the tensors are not yet computed, both values are set to \verb|-1|.

Once the exchange tensors are computed for the bond \verb|b|, they can be determined for all bonds in the orbit of \verb|b|. This should be done by calling the \verb|pull_exchange()| method for each bond in turn:
\begin{lstlisting}[style = in]
b.compute_exchange()
for gb in b.orbit:
    gb.pull_exchange()
    print(f"g_{gb.i_g} * {gb.starting_bond.exchange['DMI']} = {gb.exchange['DMI']}")
\end{lstlisting}
\begin{lstlisting}[style = out]
g_0 * [Dx, Dy, 0] = [Dx, Dy, 0]
g_1 * [Dx, Dy, 0] = [Dx, Dy, 0]
g_2 * [Dx, Dy, 0] = [-Dx, Dy, 0]
g_3 * [Dx, Dy, 0] = [-Dx, Dy, 0]
g_8 * [Dx, Dy, 0] = [Dx, Dy, 0]
g_9 * [Dx, Dy, 0] = [Dx, Dy, 0]
g_10 * [Dx, Dy, 0] = [-Dx, Dy, 0]
g_11 * [Dx, Dy, 0] = [-Dx, Dy, 0]
\end{lstlisting}
If  we use \verb|compute_exchange()| for the bonds in the orbit, the symmetry relations between different bonds will be lost. 

\subsubsection{Full data of group action}
The \verb|orbit| attribute of a bond $\vec b$ contains distinct bonds in the form $g\cdot \vec b$. If the stabilizer is non-trivial, there are several elements $g$ that map $\vec b$ to $g\cdot \vec b$, but the \verb|orbit| includes only one of them. Complete information on the group action is contained in the \verb|full_orbit| attribute of a \verb|Bond| object.  
\begin{lstlisting}[style = in]
b1 = Bond(Fe2O3).from_labels("Fe_2", [0, 0, 0], "Fe_4", [0, 0, 0])
b1.compute_exchange()

b2 = b1.orbit[2]
keys = b2.key, b2.flip().key

print(f"Symmetry elements that send b1 = {b1} \nto b2 = {b2}, up to a flip:\n")

for k in keys:
    for gb1 in b1.full_orbit[k]:
        gb1.pull_exchange()
        print(f"g_{gb1.i_g} * b1   =  {gb1} \t {gb1.exchange['DMI']}")
    print()
\end{lstlisting}
\begin{lstlisting}[style = out]
Symmetry elements that send b1 = (Fe_2, [0 0 0])--(Fe_4, [0 0 0]) 
to b2 = (Fe_6, [0 0 0])--(Fe_8, [0 0 1]), up to a flip:

g_12 * b1   =  (Fe_6, [0 0 0])--(Fe_8, [0 0 1]) 	 [0, 0, Dz]
g_14 * b1   =  (Fe_6, [0 0 0])--(Fe_8, [0 0 1]) 	 [0, 0, Dz]
g_16 * b1   =  (Fe_6, [0 0 0])--(Fe_8, [0 0 1]) 	 [0, 0, Dz]

g_18 * b1   =  (Fe_8, [0 0 1])--(Fe_6, [0 0 0]) 	 [0, 0, -Dz]
g_20 * b1   =  (Fe_8, [0 0 1])--(Fe_6, [0 0 0]) 	 [0, 0, -Dz]
g_22 * b1   =  (Fe_8, [0 0 1])--(Fe_6, [0 0 0]) 	 [0, 0, -Dz]
\end{lstlisting}
Here, the \verb|key| property of the bond \verb|b2| represents the bond as a tuple. In this form, it is used as a key in the dictionary \verb|full_orbit|. The corresponding value is the list of all bonds identical to \verb|b2| but obtained from \verb|b1| by the action of different group elements. The output indicates that the exchange tensors do not depend on the choice of the group element sending \verb|b1| to \verb|b2|. We also check that the DMI vector changes its sign upon flipping the bond, as expected. 

\subsubsection{Transformation of DM vector}
The method \verb|compute_exchange()| implements the procedure described in Sec.~\ref{Sec:SymMat} both for the symmetric exchange and for DMI. The representation matrices are constructed directly from the action \eqref{Eq:gGD} and are stored in the attribute \verb|rho| of the \verb|Bond| object. However, in the case of DMI we know that the vector $\vec D$ must transform as an axial vector, which changes sign if the bond is flipped (see Eq.~\eqref{Eq:DTransform}). Let us compare the two representations:  
\begin{lstlisting}[style = in]
b = Bond(Fe2O3).from_labels("Fe_1", [0,0,0], "Fe_11", [0,0,0])
b.compute_exchange()
rep = dict()
for i_g, flip in b.stabilizer.items():
    F = 1
    if flip:
        F = -1
    R = Fe2O3.group[i_g].R_Cart
    rep[i_g] = F * sp.det(R) * R
b.rho['DMI'] == rep
\end{lstlisting}
\begin{lstlisting}[style = out]
True
\end{lstlisting}
This illustrates that the procedure of finding representation matrices works correctly for the DMI exchange tensor. 

\subsubsection{Generating equations}\label{Sec:GenEq}
As a final example, let us generate the \LaTeX{} code for Eq.~\eqref{Eq:Ji}, which shows the full exchange matrices for a set of representative bonds in $\alpha$-Fe$_2$O$_3$:
\begin{lstlisting}[style = in]
import sympy as sp
BL = BondList(Fe2O3, "Fe", "Fe", 5, prefer_aligned= True)

equation = r"\begin{align}" + "\n"
for i_sb in range(len(BL.start_bonds)):
    
    sb = BL.start_bonds[i_sb]
    L = sb.length
    ex = sb.exchange
    
    J = ex["DMI"].matrix + ex["Symm"].matrix
    
    equation += fr"L_{{{i_sb+1}}} = {L:.2f}\text{{\AA}} \qquad " 
    equation += fr"J_{{{i_sb+1}}} &= "
    equation += sp.latex(J)

    if i_sb != len(BL.start_bonds) -1: 
        equation += r"\\" + "\n"
equation += "\n" + r"\end{align}"
\end{lstlisting}
Here, the \verb|prefer_aligned| parameter ensures that, whenever possible, the starting bond of an orbit will be chosen to belong to the coordinate planes of the Cartesian system, which can simplify the expressions of the exchange matrices.

\section{Test cases}\label{Sec:Test}

\subsection{La$_2$CuO$_4$}
We start with the famous La$_2$CuO$_4$, which is the parent compound for cuprate high-temperature superconductors. In the overdoped regime, it exhibits magnetic order with spins directed along orthorhombic $a$ axis, but canted out of the CuO$_2$ plane (see Fig. 1 in ~\cite{kastner1988}). A detailed structural study has shown that the crystal is orthorhombic in this phase, with $Cmcm$ space group~\cite{radaelli1994}.  Experimentally spins are mostly lying in the $ab$ plane and point nearly two the second nearest neighbor Cu ions, but have a small canting out of the plane.

Analysis of the exchange tensor using \textit{Jsymm} immediately reveals a non-zero Dzyaloshinskii--Moriya interaction (DMI), with the $\vec D$ vector having two independent components for the nearest-neighbor Cu--Cu bonds lying in the $ab$ plane (distance is 3.8\AA), as shown in Table~\ref{Tab:La2CuO4}. This result is in agreement with microscopic calculations of the DMI, such as those presented in \cite{shekhtman1993,Mazurenko2005}.

The presence of a $C_2$ rotational axis directed along the $c$ axis (i.e. perpendicular to each of these bonds) and passing through their centers forces the $z$ component of the DM vectors to vanish. Consequently, the DM vectors for nearest-neighbor Cu ions possess only $x$ and $y$ components, which naturally explains the canting of spins out of the $ab$ plane observed experimentally~\cite{kastner1988}.

\begin{table}[t!]
 \begin{tabular}{lccccc}
 \hline
 \hline 
$i$ & Cu & Bond vector & $\mathbf D^i$ &  $\Gamma^{i}$ \\
 \hline
1& 3-$2_1$ & (2.67, $-$2.71, 0.0)	& ($D_x$, $D_y$, 0 )  & ($\Gamma_{xx}$, $\Gamma_{yy}$, $\Gamma_{zz}$, $\Gamma_{xy}$, 0, 0)\\

2& 3-$2_2$ & ($-$2.67, 2.71, 0.0)	& ($D_x$, $D_y$, 0 ) & ($\Gamma_{xx}$, $\Gamma_{yy}$, $\Gamma_{zz}$, $\Gamma_{xy}$, 0, 0) \\

3& $2_3$-3 &($-$2.67, $-$2.71, 0.0)	& ($-D_x$, $D_y$, 0 ) & ($\Gamma_{xx}$, $\Gamma_{yy}$, $\Gamma_{zz}$, $-\Gamma_{xy}$, 0, 0) \\

4& $2_4$-3 &(2.67, 2.71, 0.0)	    & ($-D_x$, $D_y$, 0 ) & ($\Gamma_{xx}$, $\Gamma_{yy}$, $\Gamma_{zz}$, $-\Gamma_{xy}$, 0, 0)\\

 \hline
 \hline
 \end{tabular}
	\caption{Symmetry analysis of the orthorhombic structure \cite{radaelli1994} of La$_2$CuO$_4$. Bonding vectors for four nearest Cu neighbors in the $ab$ plane are given in \AA. The labels such as $3$ and $2_1$ refer to the Cu ions, as shown in Fig.~\ref{Fig:La2CuO4}. Dzyaloshinskii--Moriya vector $\mathbf D$ for each bond and components of anisotropic symmetric exchange $\Gamma$ are given in the last two columns. }
		\label{Tab:La2CuO4}
\end{table}

\begin{figure}[b!]
\centering
\includegraphics[width=0.6\textwidth]{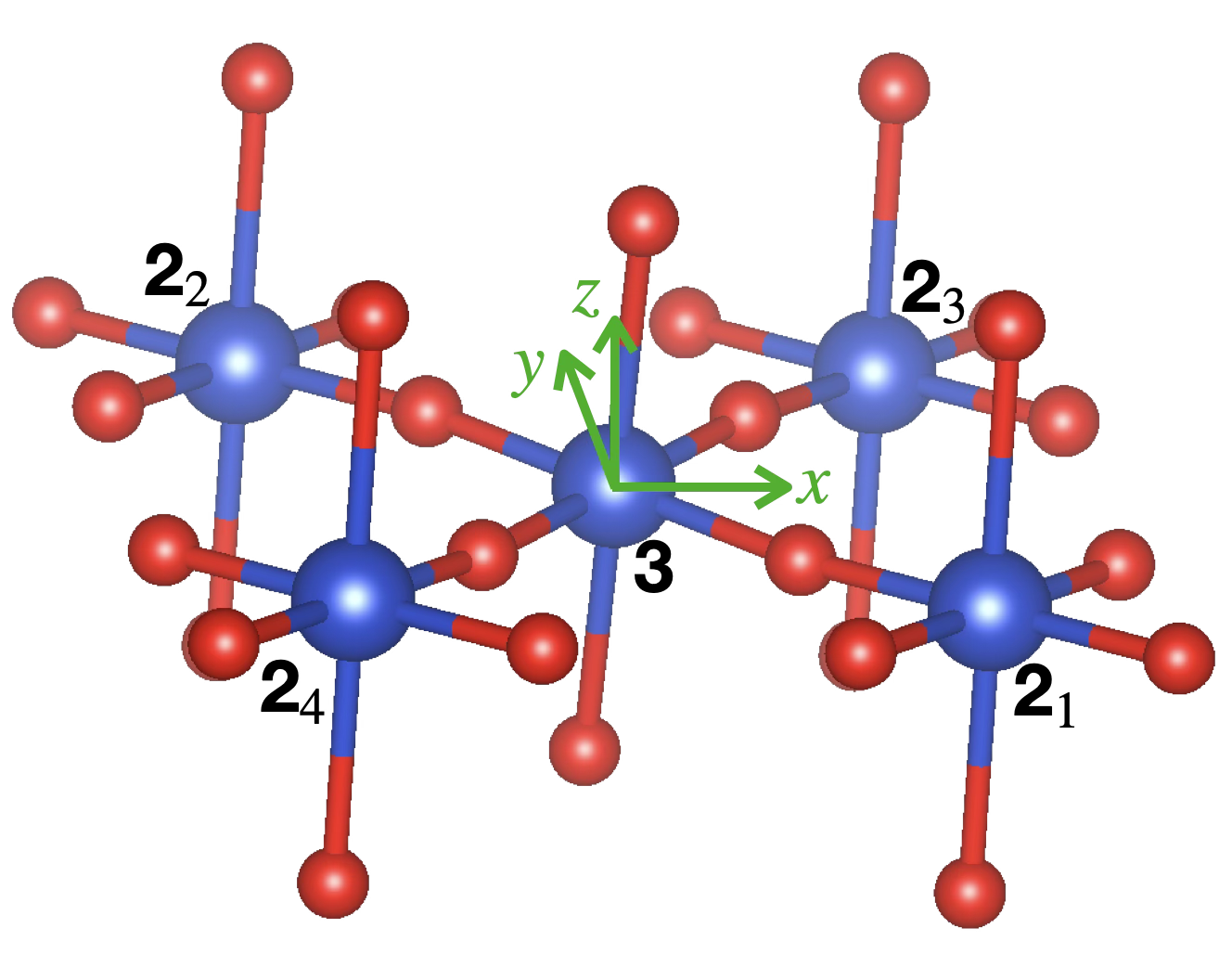}
\caption{Crystal structure of La$_2$CuO$_4$ ($ab$ plane). Cu ions are shown by blue, and O ions by red spheres. Structures were generated using VESTA \cite{Vesta}.}
\label{Fig:La2CuO4}
\end{figure}

\textit{Jsymm} also demonstrates relations between exchange tensors corresponding to different bonds that belong to the same orbit. For example, consider the DM vectors $\vec D^1$ and $\vec D^3$ of the bonds 3-$2_1$ and $2_3$-3 formed by Cu ions, as shown in Fig.~\ref{Fig:La2CuO4}. Their components are related as
\begin{equation}
    D^1_x = -D^3_x, \qquad D^1_y = D^3_y.
\end{equation}
This agrees with the results of Ref.~\cite{Mazurenko2005} (if one flips  the bond $2_3$-3, so that $\mathbf D^3$ changes its sign) and Ref.~\cite{shekhtman1993} (coordinate system must be rotated by $\tfrac{\pi}{4}$ about the $z$ axis for comparison).

\subsection{Fe$_2$O$_3$} 

Another classical example of a material with anisotropic exchange coupling is the mineral hematite, $\alpha$-Fe$_2$O$_3$, which has a corundum structure described by the $R\bar{3}c$ space group~\cite{pauling1925}. Below $T_N \sim 950$K, but above the so-called Morin temperature $T_M \sim 250$K --- the temperature of the spin reorientation transition --- it orders antiferromagnetically with moments perpendicular to the [111] direction, but with a small canting angle of $\sim 10^{-3}$ within the basal plane~\cite{flanders1965, hill2008}. Using phenomenological theory and symmetry arguments, Dzyaloshinskii was the first to explain this canting~\cite{Dzyaloshinskii1958}. This theory was later elaborated by Moriya using microscopic considerations~\cite{Moriya1960}. In this work, he also introduced the symmetry principles known as the ``Moriya rules'', which correspond in our terms to the constraints put on an individual bond by its stabilizer subgroup. 

Our analysis of the crystal structure taken from \cite{newnham1962} demonstrates that for nearest-neighbor Fe ions forming a short Fe--Fe bond (distance is 2.89\AA), the DM vector is $(0,0,D_z)$, while $\Gamma$ is allowed to have only two terms: $\Gamma_{xx}= \Gamma_{yy}$ and $\Gamma_{zz}$. DMI with a nonzero $z$ component of the DM vector causes spins lying perpendicular to the $c$ axis to cant within the (111) plane in rhombohedral notation. Exchange matrices for the next neighbors are given in Eq.~\eqref{Eq:Ji}, which was generated by the code listed in Sec.~\ref{Sec:GenEq}:
\begin{subequations}\label{Eq:Ji}
\begin{align}
L_{1} = 2.89\text{\AA} \qquad J_{1} &= \left[\begin{matrix}\Gamma_{xx} & D_z & 0\\- D_z & \Gamma_{xx} & 0\\0 & 0 & \Gamma_{zz}\end{matrix}\right]\\
L_{2} = 2.97\text{\AA} \qquad J_{2} &= \left[\begin{matrix}\Gamma_{xx} & \Gamma_{xy} & \Gamma_{xz}\\\Gamma_{xy} & \Gamma_{yy} & \Gamma_{yz}\\\Gamma_{xz} & \Gamma_{yz} & \Gamma_{zz}\end{matrix}\right]\\
L_{3} = 3.37\text{\AA} \qquad J_{3} &= \left[\begin{matrix}\Gamma_{xx} & D_z & - D_y\\- D_z & \Gamma_{yy} & \Gamma_{yz}\\D_y & \Gamma_{yz} & \Gamma_{zz}\end{matrix}\right]\\
L_{4} = 3.70\text{\AA} \qquad J_{4} &= \left[\begin{matrix}\Gamma_{xx} & D_z + \Gamma_{xy} & - D_y + \Gamma_{xz}\\- D_z + \Gamma_{xy} & \Gamma_{yy} & D_x + \Gamma_{yz}\\D_y + \Gamma_{xz} & - D_x + \Gamma_{yz} & \Gamma_{zz}\end{matrix}\right]\\
L_{5} = 3.99\text{\AA} \qquad J_{5} &= \left[\begin{matrix}\Gamma_{xx} & 0 & 0\\0 & \Gamma_{xx} & 0\\0 & 0 & \Gamma_{zz}\end{matrix}\right]
\end{align}
\end{subequations}
The degree and character of the symmetry constraints vary among the bonds. While the anisotropic symmetric exchange $\Gamma$ can be present to certain extent in all bonds, only the first, third, and fourth bonds are allowed to have DM exchange interaction. The fourth bond has the lowest possible symmetry, so its matrix has a generic form. Still, there are the symmetry restrictions for the exchange tensors of the bonds in its orbit, as we saw in Sec.~\ref{Sec:IntMode}. 

These matrices are in complete agreement with the recent results presented in Ref.~\cite{hoyer2025} (see Eq.~(25) there). The apparent discrepancy for the exchange matrices of the third bond can be removed by the rotation $U_\alpha$ of the Cartesian coordinate system by $\alpha = \tfrac{\pi}{12}$ about the $z$ axis and by selecting an appropriate bond. In our coordinates, its bond vector in Angstroms reads $[2.5173, 1.4533, 1.6957]$. Conjugation of its exchange matrix by $U_\alpha$ produces a matrix whose elements are related to each other precisely as in Ref.~\cite{hoyer2025}. This illustrates that the form of the symmetry-adapted exchange tensors may vary depending on the relative orientation of the bond vector and the axes of the Cartesian coordinate system.

\section{Summary}\label{Sec:Sum}

\textit{Jsymm} performs symmetry analysis of exchange tensors in magnetic Hamiltonians using crystallographic information from a \verb|cif| file. For any interatomic bond of interest, it computes its stabilizer subgroup and its orbit under the action of the point group of the crystal, taking into account equivalence of bonds related by lattice translations. Based on these data, \textit{Jsymm} finds the most general form of the tensors of the Dzyaloshinskii--Moriya antisymmetric exchange and anisotropic symmetric Heisenberg exchange for the selected bond, as well as the symmetry-compatible exchange tensors for all bonds in its orbit. The method uses the representation theory of finite groups: for each bond, the representation of the stabilizer on the space of exchange tensors is computed, and the invariant subspace of the trivial representation is found by projection operators. All calculations are exact, use rational and algebraic numbers via SymPy, and produce symbolic matrices, which are ready for numerical evaluations or mathematical typesetting.

The code is particularly valuable for density functional theory (DFT) calculations of magnetic materials. By determining which components of the exchange tensor are allowed by crystal symmetry, \textit{Jsymm} eliminates the need to compute all nine matrix elements of the exchange tensor for each bond. Instead, only the independent symmetry-allowed components must be evaluated, which can reduce computational time by orders of magnitude, especially when spin-orbit coupling is included. This is crucial because DFT calculations of anisotropic exchange are computationally demanding, often suffering from slow convergence and local minima. \textit{Jsymm} thus serves as an essential preprocessing tool that guides DFT calculations, ensures consistency with symmetry, and helps avoid unphysical results. The package provides both an interactive text interface and a graphical web interface, and can be used as a standalone tool or integrated into larger computational workflows.

\section{Acknowledgments}
S.V.S. thanks the Russian Science Foundation  for support via RSF 23-12-00159-P. Analisys of exchange tensors for La$_2$CuO$_4$ and Fe$_2$O$_3$ were supported by Ministry of Science and Education of Russia (via IMP).

\bibliographystyle{elsarticle-num}
\bibliography{jsymm}

\end{document}